\documentclass[preprint,aps,prd,amsmath,amssymb]{revtex4}
\usepackage{graphicx}
\usepackage{color}
\usepackage[latin1]{inputenc}
\usepackage{ulem}
\newcommand{\be}{\begin{eqnarray}}
\newcommand{\beq}{\begin{equation}}
\newcommand{\eeq}{\end{equation}}
\newcommand{\ee}{\end{eqnarray}}

\newcommand{\bmp}{\noindent\begin{minipage}{16cm}}
\newcommand{\emp}{\end{minipage}\vskip 7mm} 

\usepackage{graphicx}
\usepackage{dcolumn}
\usepackage{bm}
\usepackage{amsmath}
\usepackage{amsfonts}
\usepackage{bbm}
\usepackage{subfigure}

\def\drawbox#1#2{\hrule height#2pt
        \hbox{\vrule width#2pt height#1pt \kern#1pt
              \vrule width#2pt}
              \hrule height#2pt}

\def\Asym#1#2{\vcenter{\vbox{\drawbox{#1}{#2}
              \kern-#2pt 
              \drawbox{#1}{#2}}}}

\begin{document}
\title{Dark Majorana Particles from the Minimal Walking Technicolor}
\author{Chris {\sc Kouvaris}}
\email{kouvaris@nbi.dk}
 \affiliation{The Niels Bohr Institute,
Blegdamsvej 17, DK-2100 Copenhagen \O, Denmark, \\
University of Southern Denmark, DK-5230 Odense, and \\
CERN Theory Division, CH-1211 Geneva 23, Switzerland  }
\begin{flushright} {\rm \small CERN--PH--TH/2007--105}
\end{flushright}
\par \vskip .05in


\begin{abstract}
We investigate the possibility of a dark matter candidate emerging
from a minimal walking technicolor theory. In this case
techniquarks as well as technigluons transform under the adjoint
representation of $SU(2)$ of technicolor. It is therefore possible
to have technicolor neutral bound states between a techniquark and
a technigluon. We investigate this scenario by assuming that such
a particle can have a Majorana mass and we calculate the relic
density. We identify the parameter space where such an object can
account for the full dark matter density avoiding constraints
imposed by the CDMS and the LEP experiments.

\end{abstract}


\maketitle

\section{Introduction}

One of the most important open problems in modern physics is that
of the origin of dark matter. In 1933 Zwicky realized that the
mass from the bright part of the Coma cluster cannot explain the
motion of galaxies at the edge of the cluster. He assumed that
there must be some kind of mass, that does not interact ``much''
and therefore appears dark to us, that has to be present in order
to explain the motion of the galaxies without changing the
gravitational law. Since then it remains an enigma what is the
origin of dark matter. There are two basic types of candidates for
dark matter. In the first one belong objects usually referred as
MACHOs (Massive Compact Halo Objects), mostly of baryonic origin.
Objects like black holes, brown dwarf stars and giant planets can
be legitimate MACHO candidates. However reliable observations have
concluded that MACHOs cannot account for more than $20\%$ of dark
matter \cite{Alcock:2000ph}.

In the second type of candidates belong particles usually referred
as WIMPs (Weak Interacting Massive Particles). These particles are
usually of non-baryonic origin and in principle can account for
the whole dark matter density. There are some basic requirements
that these particles have to fulfill. First of all they have to be
electrically neutral, since in order to be part of dark matter
they should not couple to electromagnetism. In addition WIMPs
should be relatively heavy and therefore nonrelativistic, in order
to be part of cold dark matter. Very light particles (as neutrinos
for example) would form hot dark matter. The existence of hot dark
matter is not consistent with observations because the
relativistic velocities of the particles smear out structure on
small scales before the relic hot gas of light particles becomes
nonrelativistic.

There are several dark matter candidates such as axions,
supersymmetric particles and technibaryons. There are also
interesting alternative possibilities in literature
\cite{Glashow:2005jy,Fargion:2005ep,Fargion:2005xz,Khlopov:2005ew}.
Dark matter candidates are constrained theoretically as well as
experimentally. Several observations like those of WMAP give a
rather precise value for the dark matter density of the universe.
It is around $23\%$ of the total matter density. Therefore when
calculations are plausible, constraints can be put on the
different models according to what amount of dark matter they
produce. On the other hand earth based experiments like CDMS put
constraints on dark matter particles, because provided we know the
local dark matter density, the non-detection restrains the cross
section of those particles scattered off nuclei targets.

The case of dark matter candidates from technicolor theories is
not a new subject. Several authors in the past studied the
scenario of having a neutral technibaryon as a natural candidate
for dark matter
\cite{Nussinov:1985xr,Nussinov:1992he,Barr:1990ca}. Recently it
was suggested that technicolor theories that have techniquarks
transforming under not the fundamental but under higher
representations of the gauge group can be viable extensions of the
Standard Model, because they are within the limits of the
Electroweak Precision Measurements and close to the conformal
window \cite{Sannino:2004qp, Hong:2004td, Dietrich:2005jn,
Dietrich:2005wk, Sannino:2005dy, Evans:2005pu,
Gudnason:2006ug,Dietrich:2006cm}. In the minimal model only two
flavors of techniquarks and an $SU(2)$ gauge group are sufficient
to make the theory quasi-conformal. Because the addition of new
particles is small, this model is within the Electroweak
Measurements and because of the quasi-conformality this model
avoids the problems of the old technicolor theories, such as
giving mass to the heavy particles like the top quark. The
attraction of these models enhances since they can achieve
unification of couplings \cite{Gudnason:2006mk}. There can be
several different possibilities for having dark matter candidates
from these technicolor theories. A first attempt was done in
\cite{Gudnason:2006ug,Gudnason:2006yj,Accomando:2006ga}, where the
possibility of having a component of dark matter from a neutral
pseudo-Goldstone boson technibaryon was investigated. If there are
no processes violating the technibaryon number (apart from
sphalerons), and there is an initial technibaryon-antitechnibaryon
asymmetry and the neutral technibaryon is the lightest one, then
it is absolutely stable. This technibaryon  with a mass of the
order of TeV can account for even the whole dark matter density.
 However, since
in this case the WIMP is a boson, it can scatter coherently off
nuclei targets. As a result the cross section for elastic
collision with nuclei targets is four times the spin independent
one of a heavy Dirac neutrino. Such a large cross section (given
we accept that the local dark matter density in the neighborhood
of the earth is $0.2-0.4$~GeV/$\text{cm}^3$) should give a
considerable number of counts in earth based experiments like
CDMS. The CDMS collaboration has not detected any counts so far
\cite{Akerib:2004fq}. This technibaryon is ruled out as dark
matter candidate if it should account for the whole dark matter
density. However if the technibaryon consists a component of dark
matter up to $20\%$, it cannot yet be ruled out
\cite{Gudnason:2006yj}.

 Another interesting possibility of a dark
matter candidate from the same technicolor model was studied in
\cite{Kainulainen:2006wq}. The dark matter candidate in this
scenario is the neutrino of a fourth family of heavy leptons. In
the minimal walking technicolor theory with techniquarks
transforming under the 2-index symmetric representation of the
technicolor gauge group, a fourth family of leptons is needed in
order to cancel Witten global anomaly for the $SU(2)$ weak group.
If the techniquarks and the fourth family leptons have hypercharge
assignments as the corresponding Standard Model particles, then
the fourth neutrino is electrically neutral and it can account for
the whole dark matter density if the evolution in early universe
is dominated by the quintessence-like dark energy component
constrained by nucleosynthesis.

In this paper we investigate an interesting alternative
possibility to the previous scenarios. We study the case of a dark
matter candidate made of a compound bound state of a techniquark
with a technigluon forming a Majorana particle through a usual
seesaw mechanism. Because Majorana fermions cannot interact
coherently with the nucleus, such particles have smaller cross
section and therefore fewer projected counts in CDMS. We calculate
the relic density of these particles and we address the issue of
their detection. We should mention that our results are also
complementary to the scenario studied in \cite{Kainulainen:2006wq}
 as we shall explain
 in the next sections.

\section{Technicolor Model and Dark Matter Candidate}

The technicolor model we are going to use is the one used in
\cite{Gudnason:2006ug,Gudnason:2006yj,Kainulainen:2006wq}. The
technicolor group is an $SU(2)$ and there are just two
techniquarks $U$ and $D$ transforming under the adjoint
representation of $SU(2)$. The global symmetry of the model is an
$SU(4)$ that breaks spontaneously down to an $SO(4)$ resulting 9
Goldstone bosons, 3 of which are eaten by the $W$ and $Z$ bosons
\cite{Gudnason:2006ug}. The two techniquarks form a doublet under
the electroweak gauge symmetry. There are two extra particles,
i.e. a ``new neutrino'' $\nu'$ and a ``new electron'' $\zeta$
coupled to the electroweak in order to cancel the global Witten
anomaly. The authors of \cite{Gudnason:2006ug,Gudnason:2006yj}
showed that for a specific assignment of the weak hypercharge for
the technicolor particles, that is allowed by the cancellation of
gauge anomalies, one of the techniquarks (for example the $D$) is
electrically neutral. Therefore the Goldstone technibaryons of the
theory made exclusively of $D$ techniquarks, if they are the
lightest technibaryons of the theory can be a legitimate dark
matter candidate. As we mentioned in the introduction, although
such a possibility is very natural, the large cross section of the
technibaryon scattering off a nuclei target excludes this scenario
if the technibaryon consists $100\%$ of the dark matter density.

In this paper we are going to study a slightly different case. We
are going to assume the same hypercharge assignments as in
\cite{Gudnason:2006ug,Gudnason:2006yj}, so again the $D$
techniquark is electrically neutral, but we are not assuming that
the Goldstone technibaryon made of $D$ is the lightest stable
object. Rather in this scenario we assume that bound states
between $D$ techniquarks and technigluons $G$ are the lightest
objects. This is something of course not encountered  in QCD,
since it is impossible to make a colorless object out of a quark
and a gluon. This is because quarks transform under the
fundamental representation and gluons under the adjoint
representation of the gauge group. However in this particular
technicolor model both techniquarks and technigluons transform
under the adjoint representation that makes it possible to form a
colorless object. Since we have two colors, red ($r$) and green
($g$), in the adjoint representation, we have three color states:
$rr$, $rg+gr$, and $gg$. If we number these states from 1 to 3,
the objects $D_L^{\alpha}G^{\alpha}$ and $D_R^{\alpha}G^{\alpha}$
are colorless. It is assumed that we have chosen the
``appropriate'' basis for $G^{\alpha}$ and we sum over $\alpha$
which runs from 1 to 3. Apparently similar colorless states can be
constructed also using the $U$ techniquark.

Unlike in \cite{Gudnason:2006ug,Gudnason:2006yj} we assume that at
the GUT scale, Extended Technicolor (ETC) interactions violate the
technibaryon number. In addition we do not assume that there is an
initial technibaryon asymmetry. It is not necessary to speculate
regarding the particular ETC model. It is sufficient for our
purpose to assume that below the ETC scale these technibaryon
violating processes behave effectively as a Majorana mass term for
the left handed neutral techniquarks. The low energy effective
theory has mass terms of the form

\beq \dots - m_D(\psi^{\dagger}_L\psi_R + \psi^{\dagger}_R\psi_L)
-\frac{1}{2}M(\psi_L^{c\dagger} \psi_L + \psi_L^{\dagger}
\psi_L^c), \label{mass1}\eeq where $\psi_L$ and $\psi_R$ are the
left and right handed Weyl spinors of the technigluon-dressed
neutral techniquark. For example $\psi_L$ is the colorless
$D_L^{\alpha}G^{\alpha}$. The $c$ index denotes charge
conjugation, $m_D$ is the Dirac mass of the technigluon-dressed
techniquarks and $M$ is the Majorana mass for the left handed
ones. On general grounds we can give a Majorana mass also to the
right handed techniquarks or for instance we can give a Majorana
mass only to the right handed and not to the left handed
particles. Although not forbidden per se, we shall argue that the
case of left handed Majorana particles is far more interesting
from the point of view of phenomenology. The mass matrix is \beq
L_{mass}=-\frac{1}{2}\left(\psi_L^{\dagger} \psi_R^{c
\dagger}\right)\left(\begin{array}{cc}
M & m_D \\
m_D  & 0 \\
\end{array}\right)
\left(\begin{array}{c}
\psi_L^c \\
\psi_R \end{array}\right) + h.c. \label{mass2}
 \eeq

The usual seesaw mechanism gives two mass eigenvalues $M_1 =
(M+\sqrt{M^2+4m_D^2})/2$ and $M_2=(\sqrt{M^2+4m_D^2}-M)/2$ which
at the limit where $M>>m_D$ become respectively $M_1 \simeq M$ and
$M_2 \simeq m_D^2/M$. The two Majorana particles (that are mass
eigenstates) constructed from the left and right handed
techniquarks are \beq N_1 = \cos\theta \left (\begin{array}{c}
\psi_L\\
\psi_L^c\end{array}\right) + \sin\theta \left (\begin{array}{c}
\psi_R^c\\
\psi_R\end{array}\right), \label{n1}\eeq

\beq N_2 = \sin\theta \left (\begin{array}{c}
i\psi_L\\
-i\psi_L^c\end{array}\right) + \cos\theta \left (\begin{array}{c}
-i\psi_R^c\\
i\psi_R\end{array}\right), \label{n2}\eeq where the angle $\theta$
is defined through $\tan2\theta = 2m_D/M$. Varying the angle
$\theta$ within $0<\theta<\pi/4$ we can get the full range of the
ratio $m_D/M$ from zero ($m_D<<M$) to infinity ($m_D>>M$).  At the
limit where $m_D<<M$, $\tan \theta \simeq m_D/M$. Alternatively we
can write the original fields in terms of the particles $N_1$ and
$N_2$,

\beq \psi_L  = \cos\theta P_L N_1 -i \sin\theta P_L N_2,
\label{psi1}\eeq \beq \psi_R = \sin\theta P_R N_1 -i \cos\theta
P_R N_2, \label{psi2} \eeq where $P_R$ and $P_L$ are the right and
left handed projection operators $(1\pm\gamma_5)/2$. Now let's
recall how the gluon-dressed $D$ techniquark $\psi_L$ couples to
the weak gauge bosons. Since we have chosen the $D$ techniquark to
be electrically neutral, the hypercharge derived from the relation
$Q = T_3 + Y$ must be $1/2$. This means that $\psi_L$ couples only
to the $Z$ boson as \beq L_Z =
\frac{\sqrt{g^2+g'^2}}{2}Z_{\mu}\bar{\psi}_L\gamma^{\mu}\psi_L.
\label{lz}\eeq For completeness we should mention that the charge
conjugated field $\psi_L^c$ couples to the $Z$ with the same
strength but opposite sign. Now we can write how the $Z$ boson
couples to the Majorana particles $N_1$ and $N_2$. Using
Eqs.~(\ref{n1}),~(\ref{n2}),~(\ref{psi1}), and~(\ref{lz}) we get
the following couplings to the $Z$ \beq \frac{\sqrt{g^2+g'^2}}{2}
Z_{\mu}(\cos^2\theta \bar{N}_1\gamma^5\gamma^{\mu}N_1 +
\sin^2\theta \bar{N}_2\gamma^5\gamma^{\mu}N_2 + i\sin\theta
\cos\theta \bar{N}_1\gamma^5\gamma^{\mu}N_2 + h.c.) \label{N2}
\eeq It's easy to interpret the above interactions at the limit
where $m_D<<M$. Since $N_1$ is mostly $\psi_L$, it couples
strongly to the $Z$, whereas for $N_2$ being mostly $\psi_R$, the
interaction is suppressed by the factor $\sin^2\theta$. It is also
evident that the interaction among $N_1$, $N_2$ and $Z$ is
somewhat suppressed by just one power of $\sin\theta$. Because
both $N_1$ and $N_2$ are Majorana particles, the technibaryon
number is not protected as in the scenario presented in
\cite{Gudnason:2006ug,Gudnason:2006yj}. This means that two of the
$N_1$ or $N_2$ can annihilate each other. We shall show that the
heavy $N_1$ decays fast enough so its relic density today is zero.
The lighter $N_2$ is our dark matter candidate for this scenario.
We shall argue that the annihilation cross section for $N_2$ is
not big enough in order to cause the complete annihilation of its
relic density.

As we already mentioned, the $U(1)$ symmetry of the technibaryon
number is broken because of the Majorana mass term. However the
lightest technibaryon ($N_2$ in this scenario) is protected by a
$Z_2$ symmetry, i.e. the Lagrangian is invariant if $N_2
\rightarrow -N_2$. The $Z_2$ symmetry in this case is analogous to
the $R$-parity in SUSY protecting the neutralino from decaying. As
long as the ETC model respects the $Z_2$ symmetry and $N_2$ is the
lightest technibaryon, $N_2$ cannot decay, but co-annihilate with
another $N_2$.

Because of our ignorance regarding the exact ETC model and the
non-perturbative nature of the dynamics, it is difficult to
conclude decisively that a state of $DG$ can be lighter than the
regular technibaryons of the theory. However, studies of SYM with
supersymmetry softly broken showed that a Majorana mass for the
gluino $\lambda$ makes the $\lambda G$ lighter than the $\lambda
\lambda$~\cite{Evans:1997jy}. Although our model is not
supersymmetric, this is an encouraging indication that $DG$ might
be indeed the lightest technibaryon of the theory.

 By inspection of
Eqs.~(\ref{mass1}),~(\ref{mass2}),~(\ref{n1}),~(\ref{n2}),~(\ref{lz}),~and~(\ref{N2}),
one can realize that $D_LG$ couples to $Z$ with the same strength
as a left handed neutrino. In this analogy $\psi_L$ and $\psi_R$
correspond to a left and a right handed neutrino. Our scenario is
analogous to the one studied in \cite{Kainulainen:2006wq}, where
there is one left handed heavy neutrino that has either Dirac or
Majorana mass. Our study is analogous to the case where the heavy
left handed neutrino has both Majorana and Dirac mass. From this
point of view $N_1$ and $N_2$ are two Majorana neutrinos.
Therefore our results for the relic density and the CDMS and LEP
constraints are directly applicable in this case also.
\section{Relic Density of the Technicolor WIMP}

During the last few years we have obtained a lot of information
regarding the baryon and dark matter density from WMAP. The
current knowledge is that $\Omega \simeq 1$ with the baryon
density being $\Omega_Bh^2 = 0.022$ and dark matter density
$\Omega_dh^2 = 0.112$ \cite{Spergel:2006hy}. In this section of
the paper we calculate the relic density of the Majorana particle
$N_2$ and we show that it can account for the full dark matter
density for a range of masses and of the angle $\theta$. The relic
density of such a particle is governed by the well known Boltzmann
equation \beq \frac{dn_{N_2}}{dt}+3Hn_{N_2} = - \langle \sigma_A v
\rangle [ (n_{N_2})^2 - ({n_{N_2}^{eq}})^2], \eeq where $n_{N_2}$
and $n_{N_2}^{eq}$ are the number density of $N_2$ at time $t$ and
at equilibrium respectively, $H$ is the Hubble expansion rate and
$\langle \sigma_A v \rangle$ is the thermally averaged cross
section for $N_2 N_2$ annihilation times the relative velocity. On
general grounds the annihilation cross section should have the
velocity dependence $v^p$. The value $p=0$ corresponds to s-wave
annihilation and $p=2$ corresponds to a p-wave annihilation.
Indeed this is the case for Majorana particles. The thermal
velocity is $\langle v^2\rangle \sim T/m$. Therefore we can write
the annihilation cross section times the relative velocity as \beq
\langle \sigma_A v \rangle = \sigma_0 (T/m)^n = \sigma_0x^{-n},
\label{cross}\eeq where $m$ is the mass of $N_2$, $T$ is the
temperature and $x=m/T$ \cite{Kolb:1990vq}. It is understood that
the s-wave annihilation corresponds to $n=0$ and the p-wave one
corresponds to $n=1$.
 The Boltzmann equation
can be rewritten in a more convenient form in terms of
$Y=n_{N_2}/s$, ($s$ being the entropy density) as \beq
\frac{dY}{dx} = -\lambda x^{-n-2}(Y^2-Y_{eq}^2), \label{Y}\eeq
where \beq \lambda = 0.264(g_{*s}/g_*^{1/2})M_{Pl}m \sigma_0. \eeq
We define $Y_{eq} = n_{N_2}^{eq}/s$, $M_{Pl} = 1.22\times 10^{19}$
GeV. The $g_*$ and $g_{*s}$ are dimensionless numbers defined in
\cite{Kolb:1990vq}. Roughly speaking they count the total number
of effectively massless degrees of freedom. For energies above 1
 MeV, $g_*$ and $g_{*s}$ are practically identical. At a temperature
 of 1 GeV, $g_*$ and $g_{*s}$ are about 80, increasing mildly to
 roughly 100 as temperature increases up to 1 TeV.
 The knowledge of the annihilation cross section and the mass of
$N_2$ is sufficient enough to determine the relic density of
$N_2$.

 The $N_2$ couples to the $Z$ as it can be seen from
Eq.~(\ref{N2}) as a Majorana neutrino times $\sin^2 \theta$. There
are two general cases regarding the annihilation cross section of
$N_2$. The first case is when the mass of $N_2$ is smaller than
the mass of the $W$ boson and the other one when the mass is
larger. We investigate separately the two cases because different
annihilation channels contribute to each of them.

\subsection{$m<M_W$} In this case the annihilation of two $N_2$ occurs
into pairs of light fermion-antifermion (as for example light
neutrino-antineutrino pair or electron-positron pair) through $Z$
exchange. We calculated the average cross section times the
relative velocity for annihilation of two $N_2$ into a pair of
fermion-antifermion which is in accordance with \cite{Kolb:1990vq}
\beq \langle \sigma_A v\rangle = \frac{4G_F^2m^2}{3\pi}\langle
\beta^2\rangle (C_V^2+C_A^2)\sin^4 \theta, \label{little5} \eeq
where $G_F$ is the Fermi constant, and $\beta$ is the velocity of
$N_2$ at the center of mass reference system. The parameters $C_V$
and $C_A$ are defined as $C_V=j_3-2q\sin^2\theta_w$ and $C_A=j_3$,
where $j_3$ and $q$ are respectively the weak isospin and the
electric charge of the fermion and $\sin\theta_w$ is the Weinberg
angle. For the total annihilation cross section we should include
all possible channels with fermions that are lighter than $N_2$.
For a mass of $N_2$ larger than 5 GeV, the number of open channels
for annihilation into pairs of fermion-antifermion includes all
leptons and all quarks (times three colors) except the top
one~\cite{Lee:1977ua,Olive:1981ak}. The total annihilation cross
section can be written as \beq \langle \sigma_A v\rangle =
N\frac{2G_F^2m^2}{3\pi}\langle \beta^2\rangle \sin^4 \theta,
\label{little1} \eeq where $N=14.47$ represents the effective
number of channels. In principle $N$ should have been 21 since we
include five quarks times three colors and six leptons. However
since all the fermions do not couple with the same strength to the
$Z$, the total annihilation cross section is equivalent to the
total cross section of $N$ channels of neutrino-antineutrino. For
the derivation of the cross section we assumed that the fermions
are much lighter than $N_2$.
 Eq.~(\ref{little1}) is valid
only in the case where $m<<M_Z$. For larger values of $m$ we must
take into account the resonance effect and the fact that the
denominator of the propagator of the virtual $Z$ boson is not
anymore dominated by the mass of the $Z$. In this
case~(\ref{little1}) must be modified as \beq \langle \sigma_A
v\rangle = N\frac{2G_F^2m^2}{3\pi}\langle \beta^2\rangle \sin^4
\theta \frac{M_Z^4}{(s-M_Z^2)^2+\Gamma_Z^2 M_Z^2},
\label{little}\eeq where $\Gamma_Z=2.5$ GeV is the width of the
$Z$ and $s$ is the Mandelstam variable which at the
nonrelativistic limit is $s\simeq 4m^2$.
  In principle one can argue that particles like $N_2$
that couple to the $Z$ boson with a mass of a few GeV are already
excluded by constraints from the measurement of the width of the
$Z$ by the LEP collaboration. In fact, a fourth neutrino coupled
to the $Z$ with the same strength as the other three ones has been
excluded by the LEP collaboration for a mass up to 40-45 GeV
\cite{Eidelman:2004wy}. However in this case, $N_2$ can avoid
exclusion by LEP if the angle~$\theta$ is small. We can see from
Eq.~(\ref{N2}) that $N_2$ couples to the $Z$ as a regular neutrino
times $\sin^2\theta$. Therefore if $\theta$ is sufficiently small
then $N_2$ cannot be excluded by LEP even for masses smaller than
40 GeV. We address this question later on this subsection. A
similar case regarding neutrinos was studied in
\cite{Enqvist:1990yz}. Another constraint is provided by earth
based experiments for dark matter search like CDMS. However as we
shall show in the next section, the elastic cross section of $N_2$
scattering off the nuclei of the detectors is very small to be
ruled out by CDMS.

In order to calculate the relic density we have to solve
Eq.~(\ref{Y}). A very good approximate solution for
nonrelativistic particles has been given pedagogically in
\cite{Kolb:1990vq,Scherrer:1985zt}. The approximate solution for
$Y$ is \beq
 Y_{\infty}=\frac{3.79
(n+1)x_f^{n+1}}{(g_{*s}/g_*^{1/2})M_{Pl}m\sigma_0}, \eeq where
$x_f$ denotes the value of $x$ where the decoupling occurs. The
value of $x_f$ is given by the approximate relation
 \beq
 x_f \simeq \ln[(2+c)c\lambda \alpha]-(n+\frac{1}{2})\ln[\ln[(2+c)c \lambda \alpha]]. \label{x_f}\eeq The
parameter $c$ is a fitting numerical constant of order unity.
Usually the best fitting to the real solution is achieved when
$c(c+2)= n+1$. The parameter $\alpha = 0.145(g/g_{*s})$, where $g$
is the number of degrees of freedom for the particle $N_2$
(therefore $g=2$). The relic abundance is \beq \Omega_{N_2}h^2 =
Y_{\infty} s m/(\rho_{crit}/h^2)\simeq 2.82 \times 10^8 Y_{\infty}
(m/\text{GeV}). \label{omega}\eeq By inspection of~(\ref{little})
we conclude that $n=1$ and \beq \sigma_0 = \frac{NG_F^2m^2 \sin^4
\theta}{\pi}\frac{M_Z^4}{(s-M_Z^2)^2+\Gamma_Z^2 M_Z^2}.
\label{sigma0little}\eeq This is because the thermal average
velocity at the center of mass reference system is given by
 \beq \langle \beta^2 \rangle = \frac{3}{2}\frac{T}{m}.
 \label{beta}
\eeq It is easy to prove the above relation if one notices that
the thermal average velocity in the lab frame is related to the
one at the center of mass frame as $\langle\beta_{lab}^2\rangle =
2 \langle \beta^2 \rangle$. By using the equipartition theorem we
get $\langle\beta_{lab}^2 \rangle= 3(T/m)$ and therefore $\langle
\beta^2 \rangle$ is given by~(\ref{beta}). Using
Eqs.~(\ref{omega}),~(\ref{sigma0little}) and a value $g_*=100$ we
calculated the relic density of $N_2$. For $m<<M_Z$ the expression
takes the simple form \beq \Omega_{N_2}h^2 = \frac{0.0283
x_f^2}{m^2 \sin^4 \theta}. \eeq This relation is slightly more
complicated once we include the extra term of~(\ref{little})
compared to~(\ref{little1}). In Fig. 1 we show the value of
$\sin\theta$ that gives the proper relic density for $N_2$ (if it
accounts for the whole dark matter) as a function of its mass, for
a range of $m$ from 10 to 80 GeV. For a mass of 10 GeV the dark
matter density is achieved for $\sin\theta=1$. For a mass lower
than 10 GeV, $N_2$ has a relic density larger than the dark matter
density $\Omega_dh^2 = 0.112$. If we increase the mass,
$\sin\theta$ drops, reaching 0.08 for $m=45.5$ GeV, which is half
of $M_Z$. As $m$ increases beyond the resonant value, the
annihilation cross section decreases and a higher value of
$\sin\theta$ is needed in order to maintain
$\Omega_{N_2}h^2=0.112$. We have plotted $\sin\theta$ up to 80
GeV, which is the onset of a new dominant channel that we examine
in the next subsection. Since $0<\theta<\pi/4$, $\sin\theta$ is
restricted between $0<\sin\theta<\sqrt{2}/2= 0.707$. It is evident
from Fig. 1 that for $m<18$ GeV where $\sin\theta>0.707$, $N_2$
cannot provide the dark matter density and this region is
excluded. This region is also excluded by LEP as we show in the
next paragraph.

There are constraints on the masses of neutral particles that
couple to the $Z$ boson imposed by the LEP experiment. In LEP the
total decay width for the $Z$ boson into invisible neutral
particles was measured with very high accuracy. The ratio of the
decay width into invisible particles over the decay rate into a
pair of neutrino-antineutrino determines the number of light
neutral particles coupled to the $Z$. The experimental value of
this ratio is \cite{Eidelman:2004wy} \beq N_{\nu} = \frac{\Gamma(Z
\rightarrow \text{invisible})}{\Gamma(Z \rightarrow \bar{\nu}
\nu)} = 3.00 \pm 0.08. \eeq We interpret the bound as implying
that the number of light species is $N_{\nu}<3.08$. The constraint
for $N_2$ can be written as \beq 0.08 > N_{\nu}-3 = \sin^4\theta
\times \beta^3, \eeq where $\beta$ is the velocity of $N_2$
produced as $Z$ decays \cite{Ellis:1990ws}. In Fig. 1 we
implemented this constraint. As it was expected, low masses up to
23 GeV are excluded by LEP. However we can see in the figure that
LEP cannot exclude the region above 23 GeV.
 For a
typical value $m=40$ GeV, the mass of $N_1$ is $m_{N_1}\simeq 589$
GeV. This corresponds to a Majorana mass $M \simeq 549$ GeV and a
Dirac mass $m_D \simeq 153$ GeV. In the next section we shall
address the issue of $N_2$ detection by the CDMS experiment. We
shall argue that CDMS imposes no further constraints on the
suppression angle $\sin\theta$. For completeness we also checked
if it is possible for the heavier $N_1$ particle to sustain any
considerable relic density. From~(\ref{N2}) we can calculate the
decay rate of $N_1$ to an $N_2$
 and a $Z$. In order for a particle to give a
considerable relic density, the decay rate has to be smaller than
the Hubble parameter. The decay rate of $N_1$ is proportional to
\beq (\frac{\sqrt{g^2+g'^2}}{2})^2\sin^2\theta \cos^2 \theta
\frac{M_{N_1}^3}{M_Z^2}. \eeq The formula is similar to the decay
rate of the top quark. The Hubble parameter has an extremely low
value of $\sim 10^{-33}$ eV. For any realistic value of $M_{N_1}$,
and unless there is no extreme fine tuning of the mass difference
among $N_1$, $N_2$ and $Z$ or of the $\sin^2\theta \cos^2\theta$
factor, it is impossible the decay rate of $N_1$ to be smaller
than the Hubble parameter. Therefore there is no relic density for
$N_1$ since it decays very fast to $N_2$ and $Z$.
\begin{figure}[!tbp]
\begin{center}
\includegraphics[width=0.7\linewidth]{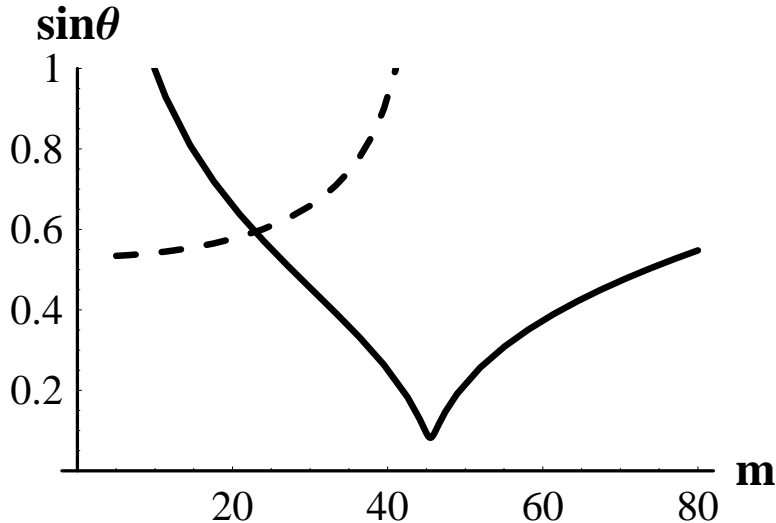}
\caption{The solid line shows the dependence of
    $\sin\theta$ on the mass of $N_2$ (in GeV), in order the relic
    density $\Omega_{N_2}h^2=0.112$. The dashed line shows the
    constraint on $m$ and $\sin\theta$ imposed by LEP. The area
    above the dashed line is excluded. This means that $m$ should be larger than
    23
    GeV, which is the value where the two curves cross each other.}
\end{center}
\end{figure}
\subsection{$m> M_W$}
The second case we investigate is the one where $m>M_W$. In
principle this means that we examine the possibility of $m$ being
higher than 80 GeV. No constraints are imposed by the LEP
experiment on this regime since the mass is higher than half of
$M_Z$. In order to calculate the relic abundance we use again the
Boltzmann Eq.~(\ref{Y}). However the annihilation cross section is
different in this case. It is very easy to show that the
annihilation of two Majorana $N_2$ into pairs of light fermions
(like electron-positron or quark-antiquark) for $m>M_Z$ is
suppressed by a factor $(1/16)(M_Z/m)^4$. This is because the
propagator of the virtual $Z$ boson is $1/(q^2-M_Z^2)$. In the
case of $m<<M_Z$ the propagator scales approximately as $1/M_Z^2$.
However if $m>>M_Z$ the propagator scales as $1/s \simeq
1/(4m^2)$. The cross section  depends on the square of the
propagator and therefore the cross section is suppressed by the
factor we mentioned above. In this regime a new channel opens up
and becomes the dominant one \cite{Enqvist:1988we}. It is the
annihilation into a pair of $W^+$-$W^-$ through a $Z$ boson. We
calculated the cross section and we found \beq \langle \sigma_A v
\rangle = \frac{G_F^2 m^2}{3 \pi} \beta^2 \beta_W
\frac{s^2}{(s-M_Z^2)^2+\Gamma_Z^2M_Z^2}\sin^4\theta (1-{\cal
O}(\frac{M_W^2}{m^2})). \eeq Again $\beta$ is the velocity of
$N_2$ at the center of mass frame and $\beta_W=\sqrt{1-4M_W^2/s}$
is the velocity of the $W$. Using~(\ref{cross}) as in the previous
case we can write $\sigma_0$ as \beq \sigma_0 = \frac{G_F^2 m^2}{2
\pi}\beta_W \frac{s^2}{(s-M_Z^2)^2+\Gamma_Z^2M_Z^2}\sin^4\theta
(1-{\cal O}(\frac{M_W^2}{m^2})). \eeq At the limit where $m>>M_W$
the above equation takes the simple form \beq \sigma_0 =
\frac{G_F^2 m^2}{2 \pi}\sin^4\theta. \eeq
 At the same limit the relic abundance of $N_2$ can be written as \beq \Omega_{N_2}h^2
= \frac{0.818 x_f^2}{m^2 \sin^4\theta}. \eeq In Fig.~2 we plot the
dependence of $\sin\theta$ as a function of the mass $m$ from 80
GeV up to 2 TeV in order to get a relic density $\Omega_{N_2}h^2
=0.112$. In our plot we took into account both the annihilation
channel to $W^+$-$W^-$ and to pairs of fermions-antifermions. For
the $W^+$-$W^-$ channel we dropped the terms that scale as powers
of $(M_W/m)^2$. The mixing angle $\sin\theta$ has a peak at 122
GeV and then it drops smoothly as $m$ increases. It is easy to see
why this peak appears. As soon as $m$ becomes larger than 80 GeV,
it is possible to have annihilation to a pair of $W^+$-$W^-$.
However close to the onset, the phase space for this amplitude is
very small and the cross section is controlled by $\beta_W$ (which
is zero at $s=4M_W^2$). Between 80 and 122 GeV, the total
annihilation cross section drops because the $W^+$-$W^-$ channel
has not yet enough phase space and the fermion-antifermion
channels that still dominate have a cross section that falls as we
explained at the beginning of this subsection. Once $m$ becomes
large enough so there is a lot of phase space for the $W^+$-$W^-$
annihilation, the cross section increases. This means that
$\sin\theta$ must drop if we have to maintain the dark matter
density.
 At a
mass of 1 TeV $\sin\theta = 0.26$. For this mass of $N_2$, the
corresponding mass for the heavy $N_1$ is 13.5 TeV and the
original Dirac and Majorana masses are respectively $m_D=3.7$ TeV
 and $M=12.5$ TeV.
 We plot $\sin\theta$ up to $m=2$ TeV where
$\sin\theta=0.19$. Our calculation of the total cross section and
consequently of the value of $\sin\theta$ is extremely accurate
both at the onset of the $W^+$-$W^-$ channel and for $m>>M_W$. The
only region where the cross section is not very accurate is at the
peak (around 122 GeV) because this is where the corrections of the
order of $M_W^2/m^2$ are important. For masses larger than 122
GeV, these corrections are suppressed. For masses close to the
onset of the $W^+$-$W^-$ channel, these corrections are
unimportant because the annihilation cross section is still
dominated by the annihilation to pairs of fermions-antifermions.
However even at the peak, our estimation for the $\sin\theta$ is
off at most by $\sim 10\%$. Either on the left or on the right of
the peak our estimation of $\sin\theta$ becomes better than $95\%$
accurate within a few GeV.

\begin{figure}[!tbp]
\begin{center}
\includegraphics[width=0.7\linewidth]{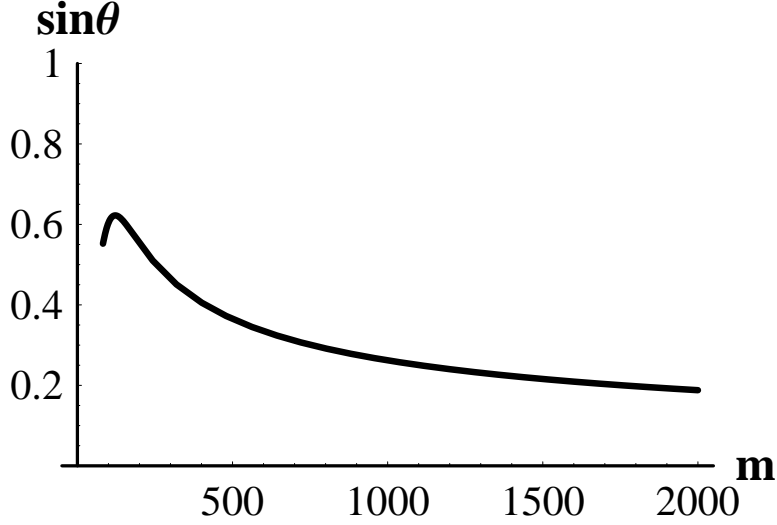}
\caption{As in Fig.~1 the solid line shows the dependence of
    $\sin\theta$ on the mass of $N_2$ (in GeV), in order the relic
    density $\Omega_{N_2}h^2=0.112$. }
\end{center}
\end{figure}

\section{Detection of the Lightest Technibaryon in CDMS}

We turn now our attention to the question of detection of $N_2$
from dark matter search experiments. It is well known that earth
based experiments like CDMS can put tight constraints regarding
the cross section of WIMPs scattering off nuclei targets. In fact,
the most important constraint related to the scenario of
techni-Goldstone boson dark matter candidate was coming from the
non-detection of counts in CDMS
\cite{Gudnason:2006ug,Gudnason:2006yj}. There are two basic
factors that influence the number of counts on earth detectors.
The first one is the local dark matter density and the second one
is the elastic scattering cross section between the WIMP and the
nuclei of the detector. Most cosmologists agree that the local
dark matter density should be somewhere between $0.2-0.4$
GeV/$\text{cm}^3$. As for the cross section, Majorana particles
have usually much smaller cross section compared to Dirac ones
because Majorana fermions do not scatter coherently with the whole
nucleus of the target. This is our motivation for investigating
$N_2$ as a dark matter candidate. A review of the cross section of
different dark matter candidates can be found in
\cite{Vergados:2006ry,Oikonomou:2006mh}. For a Majorana particle
only spin-dependent elastic collisions contribute
\cite{Lewin:1995rx,Goodman:1984dc}. Following \cite{Lewin:1995rx},
we can write the spin-dependent cross section for $N_2$ as \beq
\sigma_{N_2} = \frac{2G_F^2}{\pi}\mu^2 I_s \sin^4\theta,
\label{crossn2} \eeq
 where
 $\mu$ is the reduced mass of the system WIMP-nucleus and $I_s$
is conventionally written in the form $I_s = C^2\lambda^2J(J+1)$.
$C$ is given by \beq C=\sum_q T_q^3 \Delta_q \qquad (q=u,d,s),
\eeq where $\Delta_q$ is the fraction of the spin carried by the
specific quark $q$. $T_q^3$ is the 3rd component of the isotopic
spin of each of the three quarks
$(T_u^3=1/2,T_d^3=-1/2,T_s^3=-1/2)$. The values for the different
$\Delta_q$ given by the European Muon Collaboration (EMC) are
$\Delta_u =0.83$, $\Delta_d=-0.43$ and $\Delta_s=-0.10$
\cite{Lewin:1995rx}. A realistic value for $\lambda^2J(J+1)$
within the model of odd group for the detectors of
$\text{Ge}^{73}$ is 0.065. Given these values, the overall factor
$I_s \simeq 0.03$ for the Ge detectors. The cross section can be
written in convenient units pb as \beq \sigma_{N_2} = 3.38 \times
10^{-2}\mu^2 I_s \sin^4\theta= 1.01 \times 10^{-3} \mu^2
\sin^4\theta (pb). \eeq The total rate of counts on an earth based
detector in experiments like CDMS is \cite{Lewin:1995rx} \beq
R_0=\frac{540}{A m}\left(\frac{\sigma_0}{1{\rm
 pb}}\right)\left(\frac{\rho_{dm}}{0.4{\rm GeVc}^{-2}{\rm
 cm}^{-3}}\right)\left(\frac{\upsilon_0}{230{\rm kms}^{-1}}\right){\rm
 kg}^{-1}{\rm days}^{-1},
\eeq where $A$ is the mass number of the nucleus of the detector,
$\rho_{dm}$ is the local dark matter density and $\upsilon_0$ is
the average velocity of the WIMP. The total rate is given in terms
of $\text{kg}^{-1}\text{days}^{-1}$ which means that for a given
detector of
 mass $x$ and of exposure time $y$, the total rate must be multiplied
 by $xy$. However the number of actual counts that can be seen in a
 detector is given by \beq  {\rm counts} =
\frac{dR}{dT}\Delta T \times \tau \ , \label{rate4} \eeq where
$\tau$ is the exposure of the detector measured in
$\text{kg}\cdot\text{days}$ and $\Delta T$ is the energy
resolution of the detector. The factor $dR/dT$ is the derivative
of the total rate with respect to the recoil energy $T$ given by
the approximate relation \beq
  \label{rate1}
\frac{dR}{dT}=c_1\frac{R_0}{E_0 r}e^{-c_2T/E_0r} ,
 \eeq
where $E_0$ is the kinetic energy of the WIMP and
$r=4mM_n/(m+M_n)^2$, $m$ and $M_n$ being the masses of the WIMP
and the nucleus of the detector respectively. The $c_1$ and $c_2$
are fitting parameters. Eq.~(\ref{rate1}) was derived in
\cite{Lewin:1995rx} after averaging over the Boltzmann velocity
distribution of the WIMP. The case with $c_1=c_2=1$ corresponds to
averaging of the velocity from zero to infinity. However it has
been pointed out that the motion of the earth should be taken into
account and more realistic values for the parameters are
$c_1=0.751$ and $c_2=0.561$. These parameters depend mildly on the
detector's energy threshold and the mass of the WIMP, however do
not change a lot and we consider them as constants. We have taken
the velocity of the earth to be $v_E = 1.05 \times v_0 = 1.05
\times 230 \text{km/sec}$. In the first results of the CDMS
experiment \cite{Akerib:2004fq}, the exposure of the Ge detectors
was 19.4 $\text{kg}\cdot\text{days}$. The energy resolution
$\Delta T= 1.5$ keV and the recoil energy threshold is 20 keV
although the detector can count recoil energies down to 10 keV.
The current exposure of the detectors in CDMS (19.4
$\text{kg}\cdot\text{days}$) is not sufficient to give any counts
for a particle like $N_2$ with local dark matter density ranging
between $0.2-0.4$~GeV/$\text{cm}^3$. This is true for the whole
range of $m$ we examined. In Fig.~3 we show what is the required
exposure in order to detect one count of $N_2$ with $90\%$
confidence level as a function of $m$. The $90\%$ confidence level
corresponds to 2.3 counts. For the first case we studied with $m$
up to 80 GeV, the required exposure increases as a function of $m$
up to the resonance peak of $m=45.5$ GeV and then drops. For a
local dark matter density $\rho =0.4$ GeV$/\text{cm}^3$, a total
amount of 7004 $\text{kg}\cdot\text{days}$ is needed for a typical
mass $m=30$ GeV. This practically means that the required exposure
for detection of a single count should be 361 times the current
exposure of 19.4 $\text{kg}\cdot\text{days}$.
 For masses $m>>100$ GeV, the required factor is much larger.
  For $m=2$ TeV, the required exposure reaches $8 \times 10^6$ $\text{kg}\cdot\text{days}$.
Our results are in accordance with the predictions of the CDMS
group for Majorana dark particles \cite{Akerib:2005za}.
\begin{figure}[!tbp]
  \begin{center}
    \mbox{
      \subfigure{\resizebox{!}{4.8cm}{\includegraphics{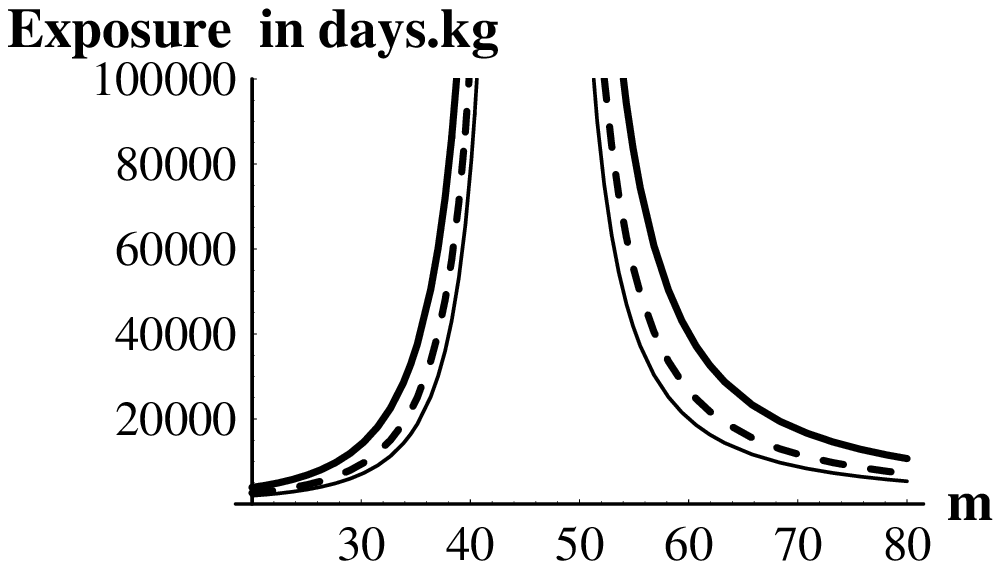}}} \quad
      \subfigure{\resizebox{!}{4.8cm}{\includegraphics{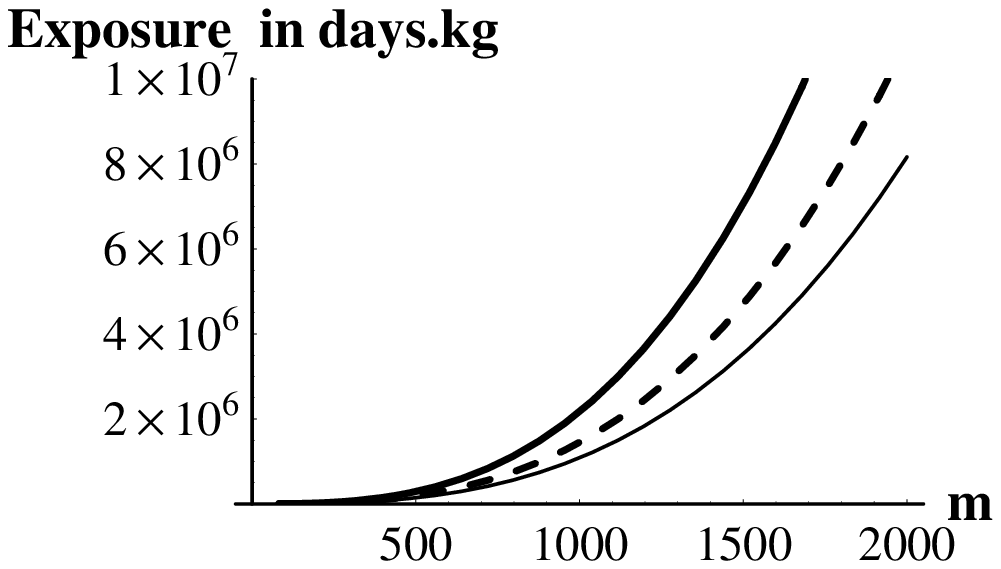}}}
      }
    \caption{\emph{Left Panel}:The required exposure of the
    Ge detectors in $\text{kg}\cdot\text{days}$ for a single count
    (with 90$\%$ confidence level) as a function of $m$ (in GeV) for the range
    $20<m<80$, although in reality $m$ is constrained by LEP to be larger than 23 GeV.
     The thin solid line corresponds to local dark
    matter density $\rho=0.4 ~\text{GeV}/\text{cm}^3$, the dashed
    one to $\rho=0.3 ~\text{GeV}/\text{cm}^3$ and the thick solid
    one to $\rho=0.2 ~\text{GeV}/\text{cm}^3$. For the purposes of presentation we show
    the required exposure up to 100000 $\text{kg}\cdot
    \text{days}$. Around the resonance, where $m=45.5$ GeV, the
    required exposure has a sharp peak of about $10^7$ $\text{kg}\cdot
    \text{days}$.
     \emph{Right Panel}:As in the left panel for $80<m<2000$ GeV.}
    \label{fig:Omega}
    \end{center}
\end{figure}
\section{The Case of Majorana Mass for The Right Handed Particle}

 So far we discussed the case of a Majorana mass for the
left handed $D_LG$ and a Dirac mass for both $D_LG$ and $D_RG$.
However, one might ask the question of what happens if instead of
giving a Majorana mass to the left handed particle, we give it to
the right handed one ($D_RG$). This means that in the mass matrix
of Eq.~(\ref{mass2}), $M$ and 0 in the diagonal are exchanged. It
turns out that the two Majorana eigenstates can be described
easily using Eqs.~(\ref{n1}) and~(\ref{n2}) if we make the
substitution $\sin\theta \rightarrow \cos\theta$ and $\cos\theta
\rightarrow \sin\theta$. Under this description, the relation that
defines the angle $\theta$, $\tan2\theta=2m_D/M$ remains the same.
In addition, the ratio of the masses of $N_1$ and $N_2$ is given
by \beq \frac{M_{N_2}}{M_{N_1}}=\tan^2\theta, \label{m1m2}\eeq as
in the previous case. Using this description, $N_1$ couples to the
$Z$ with a factor $\sin^2\theta$ and $N_2$ with a factor
$\cos^2\theta$. Since $N_2$ is lighter than $N_1$, we can find the
value of $\cos\theta$ as a function of $M_{N_2}$ in order to have
$\Omega_{N_2}h^2=0.112$. This is exactly what we did in the
previous sections apart from the fact that the annihilation cross
section for $N_2$ is not proportional now to $\sin^4\theta$, but
$\cos^4\theta$. This means that Figs. 1 and 2 are valid in this
case if we substitute in the vertical axis of the figures
$\sin\theta$ by $\cos\theta$. However because~(\ref{m1m2}) remains
unchanged, in order for $N_2$ to be lighter than $N_1$,
$\sin\theta<\cos\theta$. This happens when
$\cos\theta>\sqrt{2}/2\simeq 0.707$. By inspection of Figs. 1 and
2, one can see that there is only one region where this is true.
It is below 18 GeV (as seen in Fig. 1) and it is already excluded
by LEP.

The physical reason of the qualitative difference between the two
general cases we studied, namely giving a Majorana mass to either
the left or the right handed particles relies on the simple fact
that in the first case the lighter Majorana is also the particle
with the suppressed annihilation cross section and therefore the
one that can provide a considerable abundance. In the second case,
the lighter Majorana is the one that is ``mostly'' left handed and
therefore the big annihilation cross section cannot make this
particle to sustain a substantial relic density.

    \section{Conclusions and Discussion}
    In this paper we investigated the possibility of a dark matter
    candidate emerging from the minimal walking technicolor
    theory. Because the two techniquarks of the theory transform
    under the adjoint representation of the technicolor $SU(2)$
    group, it is possible to have a bound colorless state between
    a techniquark and a technigluon. We looked upon the scenario
    that the left handed technigluon-dressed techniquark has a
    Majorana mass and both left and right handed have a Dirac
    mass. We found that this dark matter candidate can account
    for the whole dark matter density for practically any mass
    higher than 23 GeV.
     This dark matter candidate can account for the
    whole dark matter density without being ruled out by LEP or
    CDMS.
    We also commented on what happens if it is the right handed particles
    that have a Majorana mass instead of the left handed. We showed that in
    this case the lighter Majorana particle cannot account for
    the whole dark matter density.
     Since we are lacking the tools to
    calculate the spectrum of this technicolor theory and we don't know the exact ETC model,
     we cannot know a priori what is the mass of $N_2$. Lattice
     methods just started being implemented for studying the dynamics of models with fermions
     in higher
     representations of the gauge group than just fundamental. It
     will be very interesting if it will be possible to study in
     lattice this bound state of quark-gluon.

     We should emphasize here that our results are complementary
     to
     the case studied in \cite{Kainulainen:2006wq}. We already
     mentioned that in the minimal walking technicolor model with
     techniquarks in the adjoint representation, it is necessary
     to have an extra family of leptons to cancel Witten global
     anomaly. If the hypercharge assignment for the fourth
     neutrino is like in the Standard Model, then this heavy
     fourth neutrino can play the role of a dark particle. However
     in the candidate we studied, we use a different hypercharge
     assignment, the one that makes $D$ neutral. Both assignments
     are consistent and free of gauge anomalies. Although the
     hypercharge assignments are different, the strength of how
     $D_L G$ couples to the $Z$ boson is the same as this of the
     fourth neutral neutrino. Therefore if one assumes that the fourth
     left handed neutrino (coming from technicolor) has both Majorana
    and  Dirac mass, then the calculation of the relic
     density and the constraints from LEP and CDMS are
     identical with the corresponding ones of $N_2$.

 \acknowledgments
 I would like to thank F. Sannino for reading carefully the manuscript and for
 discussions. I am also grateful to K. Tuominen, and professors J.D.
 Vergados and M. Khlopov for their comments.

The work of C.K. is supported by the Marie Curie Excellence Grant
under contract MEXT-CT-2004-013510.

\end{document}